\documentclass[12pt]{article}
\usepackage{latexsym}
\usepackage{amssymb}
\usepackage{relsize}
\usepackage{geometry}
\usepackage{tensor}
\usepackage{csquotes} 
\usepackage{graphicx}
\graphicspath{ {img/} }

\usepackage{showlabels}

\def\beqn{\begin{eqnarray}}
\def\eeqn{\end{eqnarray}}

\def\beq{\begin{equation}}
\def\eeq{\end{equation}}
\def\ba{\beq\new\begin{array}{c}}
\def\ea{\end{array}\eeq}

\def\Tr{{\rm Tr}}
\newcommand{\gsim}{\lower.7ex\hbox{$
\;\stackrel{\textstyle>}{\sim}\;$}}
\newcommand{\lsim}{\lower.7ex\hbox{$
\;\stackrel{\textstyle<}{\sim}\;$}}

\newcommand{\ntwo}{${\mathcal N}=2$ }

\newcommand{\none}{${\mathcal N}=1$ }

\newcommand{\vp}{\varphi}
\newcommand{\pt}{\partial}

\newcommand{\p}{\partial}
\newcommand{\wt}{\widetilde}
\newcommand{\ov}{\overline}
\newcommand{\mc}[1]{\mathcal{#1}}

\newcommand{\lgr}{\left\lgroup}
\newcommand{\rgr}{\right\rgroup}

\newcommand{\ue}{{\rm U}(1)}
\def\slashed#1{\setbox0=\hbox{$#1$}             
   \dimen0=\wd0                                 
   \setbox1=\hbox{/} \dimen1=\wd1               
   \ifdim\dimen0>\dimen1                        
      \rlap{\hbox to \dimen0{\hfil/\hfil}}      
      #1                                        
   \else                                        
      \rlap{\hbox to \dimen1{\hfil$#1$\hfil}}   
      /                                         
   \fi}                                        %

\newcommand{\sun}{{\rm SU(}N{\rm )}}

\newcommand{\mUp}{m_{\rm U(1)}^{+}}
\newcommand{\mUm}{m_{\rm U(1)}^{-}}
\newcommand{\mNp}{m_\text{SU($N$)}^{+}}
\newcommand{\mNm}{m_\text{SU($N$)}^{-}}
\newcommand{\AU}{\mc{A}^{\rm U(1)}}
\newcommand{\AN}{\mc{A}^\text{SU($N$)}}
\newcommand{\aU}{a^{\rm U(1)}}
\newcommand{\aN}{a^\text{SU($N$)}}

\newcommand{\nbar}{\ov{n}}

\usepackage{mathtools}
\usepackage{hyperref}
\usepackage{xcolor}

\usepackage{listings}
\definecolor{mygreen}{RGB}{28,172,0} 
\definecolor{mylilas}{RGB}{170,55,241}

\begin{document}

\hypersetup{%
	linkbordercolor=blue,
}

%
%


\begin{titlepage}


\begin{center}
\Large{{\bf What Becomes  of Semilocal non-Abelian Strings in \none Supersymmetric QCD}}

\vspace{5mm}

{\large \bf E.~Ievlev$^{\,a,b}$ and  A.~Yung$^{\,a,b,c}$}
\end {center}

\begin{center}

$^{a}${\it National Research Center ``Kurchatov Institute'',
Petersburg Nuclear Physics Institute, Gatchina, St. Petersburg
188300, Russia}\\
$^{b}${\it  St. Petersburg State University,
 Universitetskaya nab., St. Petersburg 199034, Russia}\\
 $^c${\it  William I. Fine Theoretical Physics Institute,
University of Minnesota,
Minneapolis, MN 55455}\\
\end{center}

\vspace{1cm}

\begin{abstract}

We study non-Abelian vortex strings in \ntwo supersymmetric QCD with the gauge group U$(N)$ deformed by 
the mass $\mu$ of the adjoint 
matter. This deformation breaks \ntwo supersymmetry down to \none and in the limit of large $\mu$ the theory
flows to \none QCD. Non-Abelian strings in addition to translational zero modes  have orientation moduli.
In the \ntwo limit of small $\mu$ the dynamics of orientational moduli is described by the  two dimensional $CP(N-1)$ model  for QCD with $N_f=N$ flavors of quark hypermultiplets. For the case of $N_f>N$ the non-Abelian string becomes semilocal developing additional size moduli which modify the effective two dimensional $\sigma$-model on the string making its target space non-compact. In this paper we consider the $\mu$-deformed theory with $N_f>N$  eventually   making $\mu$ large. We show that size moduli develop a potential that forces the string transverse size to shrink. Eventually in the large $\mu$ limit size moduli decouple and the effective theory on the string
reduces to the $CP(N-1)$ model. We also comment on physics of confined monopoles.

\end{abstract}

\end{titlepage}

\newpage


%
%

\section*{Introduction}  \label{introduction}
\addcontentsline{toc}{section}{Introduction}

Searches for a  non-Abelian generalization of Seiberg-Witten scenario of quark  confinement \cite{SW1,SW2} 
leads to the  
discovery of  non-Abelian vortex strings  in  \ntwo supersymmetric QCD \cite{HT1,ABEKY,SYmon,HT2}, see also \cite{Trev,Jrev,SYrev,Trev2} for  reviews. They are formed in the Higgs phase of U$(N)$ gauge theory due to the condensation of  scalar quarks and are responsible for the confinement of monopoles. In the strong coupling
regime the theory is in the \textquote{instead-of-confinement} phase. Particularly rich structure with 
non-Abelian dual gauge group appears in a theory with number of quark flavors $N_f >N$, see \cite{SYdualrev}
for a review.

The \ntwo SQCD is a nice theoretical laboratory to study non-perturbative non-Abelian dynamics. However, since we wish to learn more about the \textquote{real world}, we are interested to study more realistic theories. \none
supersymmetric QCD is one of the most promising examples. Much in the same way as the real world QCD it has no adjoint scalars and no Abelianization of the theory can occurs due to their condensation.

To study this theory we start with \ntwo SQCD deformed by the mass  $\mu$  of  the adjoint matter.
This deformation breaks \ntwo supersymmetry down to \none. In the limit of large $\mu$ the adjoint matter decouples and the theory flows to \none QCD. It was shown that the non-Abelian \textquote{instead-of-confinement} phase survives for $N_f >N$, see review \cite{SYdualrev} and references therein. 

Motivated by these results  in this paper we study non-Abelian confining strings in $\mu$-deformed 
\ntwo SQCD with $N_f>N$. In particular, we consider the limit of large $\mu$ when the theory flows to \none 
SQCD.

The  case $N_f=N$ was studied earlier. To the leading order in $\mu$ the mass term for the adjoint matter 
reduces to Fayet-Iliopoulos (FI) $F$-term which does not break \ntwo supersymmetry \cite{Hanany:1997hr,VY}.
In the quark vacuum squark condensate is determined by $\sqrt{\mu m}$, where $m$ is a quark mass. 
In this setup non-Abelian strings were first found \cite{HT1,ABEKY,SYmon,HT2} and 
their dynamics was well studied, see \cite{SYrev} for a review. 
 In addition to the
translational zero modes typical for Abelian Abrikosov-Nielsen-Olesen (ANO) vortex strings  \cite{ANO}, 
non-Abelian strings have  orientational moduli associated with rotations of their fluxes inside the non-Abelian SU$(N)$ group. The dynamics of the orientational moduli in \ntwo QCD is described by the two dimensional 
CP$(N-1)$ model living on the world sheet of the non-Abelian string.

The $\mu$ deformation of \ntwo SQCD was considered recently in \cite{YIevlevN=1}. It turns out that
 the non-Abelian string ceases to be BPS, and world sheet supersymmetry is completely lost. Fermionic sector of the low energy world sheet  theory decouples at large $\mu$, while the bosonic sector is given by two dimensional  CP$(N-1)$ model. It was also shown that in the case of equal quark masses confined monopoles seen in the world sheet theory as kinks \cite{SYmon,HT2} survive $\mu$ deformation and present in the limit of \none 
SQCD. The  potential in  two dimensional world sheet theory induced by quark mass differences was also found. 

Non-Abelian strings in \ntwo
SQCD with \textquote{extra} quark flavors  ($N_f > N$) were also well studied. In this setting the  string 
 develops size moduli and becomes semilocal. In particular, in the Abelian case these strings interpolate between ANO local strings and sigma-model lumps \cite{Vachaspati:1991dz,Hindmarsh:1991jq,Hindmarsh:1992yy,Preskill:1992bf,AchVas}. World-sheet theory on the semilocal non-Abelian string  was first 
considered from a D-brane prospective \cite{HT1,HT2}, and later from a field theory side \cite{Shifman:2006kd,Eto:2006uw,Eto:2007yv,ShYVinci}. In particularly, in \cite{ShYVinci}  
it was found that the world sheet theory is the so-called  ${\mathcal N}= (2,2)$ supersymmetric $zn$ model.

In this paper we continue studies of non-Abelian strings in SQCD with additional quark flavors, $N_f>N$
and consider $\mu$ deformed theory. In particular, we study what becomes of semilocal non-Abelian strings
as we increase $\mu$ and take the large $\mu$ limit where the theory flows to \none SQCD. First we found that much in the same way as for $N_f=N$ case \cite{YIevlevN=1} the string is no longer  BPS and the world sheet supersymmetry is lost. 

Moreover, 
 as we switch on the deformation parameter $\mu$ the string itself ceases to be semilocal. Considering the world sheet theory at small $\mu$ we show that string  size moduli develop a potential which forces them to shrink. Eventually in the large $\mu$ limit size moduli decouple and the effective theory on the string
reduces to CP$(N-1)$ model. 

We also briefly discuss the  physics of confined monopoles.

The paper is organized as follows. In Sec. \ref{sec:basics} we briefly outline the underlying bulk theory and calculate its mass spectrum. In Sec. \ref{sec:string-small_mu} we consider the non-Abelian semilocal string in the  $\mu$ deformed theory and  study  the world sheet theory for this string. 
We summarize our results in Sec. \ref{sec:results}.

%
%

\section{Theoretical setup \label{sec:basics}}

\subsection{Bulk theory}

In this section we briefly describe our initial theory in the bulk. The basic model is four-dimensional \ntwo supersymmetric QCD with the gauge group SU$(N)\times$U(1). 
The field content of the theory is as follows. The matter consists of $N_f = N + \tilde{N}$ flavors of quark hypermultiplets in the fundamental representation, scalar components being $q^{kA}$ and $\wt{q}_{Ak}$.  Here, $A = 1,..,N_f$ is the flavor index and $k=1,..,N$ is the color index.
The vector multiplets consist of U(1) gauge field $A_\mu$ and SU($N$) gauge field $A^a_\mu$, complex scalar fields $a$ and $a^a$ in the adjoint representation of the color group, and their Weyl fermion superpartners. Index $a$ runs from 1 to $N^2 - 1$, and the spinorial index $\alpha = 1,2$. 

Superpotential of the \ntwo SQCD is 
\begin{equation}
 \mc{W}_{\mc{N}=2} ~~=~~  \sqrt{2}\, \left\{ 
	 \frac{1}{2}\wt{q}{}_A \AU q^A ~+~
	 \wt{q}{}_A \mc{A}^a T^a q^A \right\}  ~+~
	 m_A\, \wt{q}{}_A q^A \ ,
\label{ntwo_superpotential_general}
\end{equation}
which includes  adjoint matter chiral \none multiplets $ \AU $ and $ \AN = \mc{A}^a T^a $, and the quark chiral \none multiplets $ q^A $ and $ \wt{q}{}_A $ (here we use the same notation for the quark superfields and their scalar components). The $\mu$ deformation considered in this paper is given by the superpotential
\begin{equation}
 \mc{W}_{\mc{N}=1} ~~=~~ \sqrt{\frac{N}{2}}\,\frac{\mu_1}{2} \left(\mc{A}^{\rm U(1)}\right)^2  ~~+~~
	 \frac{\mu_2}{2} \left( \mc{A}^a \right)^2 ~.
\label{none_superpotential_general}
\end{equation}
We assume the deformation parameters to be of the same order, $\mu_1 \sim \mu_2 \sim \mu$. When we increase $\mu\to\infty$, \ntwo supersymmetry becomes broken, and the theory flows to \none SQCD. 
Instead in the limit of small $\mu$ this superpotential does not break the \ntwo supersymmetry and reduces to a Fayet--Iliopoulos (FI) $F$-term \cite{Hanany:1997hr, VY}. 

In order to control  the theory and stay at weak coupling as we take this limit, we require the product $\sqrt{\mu m}$ to stay fixed and well above $\Lambda_{{\cal N}=1}$, which is the scale of the \sun~sector of \none QCD .

The bosonic part of the action is given by
%
\begin{align}
\label{theory}
S_{\rm bos} ~~=~~ & \int d^4 x 
\lgr
\frac{1}{2g_2^2}\Tr \left(F_{\mu\nu}^\text{SU($N$)}\right)^2  ~+~
\frac{1}{4 g_1^2} \left(F_{\mu\nu}^{\rm U(1)}\right)^2 ~+~ 
\right. 
\\[2mm]
\notag
&
\phantom{int d^4 x \lgr\right.}
\frac{2}{g_2^2}\Tr \left|\nabla_\mu \aN \right|^2   ~+~
\frac{1}{g_1^2} \left|\p_\mu \aU \right|^2
~+~
\left| \nabla_\mu q^A \right|^2 ~+~ \left|\nabla_\mu \ov{\wt{q}}{}^A \right|^2 
~+~
\\[2mm]
\notag
&
\phantom{int d^4 x \lgr\right.}
\left.
V(q^A, \wt{q}{}_A, \aN, \aU)
\rgr .
\end{align}
Here $ \nabla_\mu $ is the covariant derivative in the corresponding representation:
\begin{align*}
\nabla_\mu^{\rm adj} & ~~=~~ \p_\mu  ~-~ i\, [ A_\mu^a T^a, \;\cdot\;]~, \\[3mm]
\nabla_\mu^{\rm fund} & ~~=~~ \p_\mu ~-~ \frac{i}{2} \,A^{\rm U(1)}_\mu ~-~ i\, A_\mu^a T^a~,
\end{align*}
with the SU($N$) generators normalized as
$\Tr \left( T^a T^b \right) ~=~ (1/2) \, \delta^{ab}~$.
%
Superpotentials \eqref{ntwo_superpotential_general}, \eqref{none_superpotential_general} contribute to the scalar potential $V$ which is given by the sum of $F$ and $D$ terms, 
\begin{align}
\notag
& V(q^A, \wt{q}{}_A, \aN, \aU) ~~=~~ 
\\[3mm]
\notag
&\qquad\quad ~~=~~
\frac{g_2^2}{2} \left( \frac{1}{g_2^2}\,f^{abc}\ov{a}{}^b a^c 
~+~ \ov{q}{}_A\, T^a q^A ~-~ \wt{q}{}_A\, T^a \ov{\wt{q}}{}^A \right)^2 
\\[3mm]
\label{V}
&\qquad\quad ~~+~~
\frac{g_1^2}{8}\, (\ov{q}{}_A q^A ~-~ \wt{q}{}_A \ov{\wt{q}}{}^A)^2
\\[3mm]
\notag
&\qquad\quad ~~+~~
2g_2^2\, \Bigl| \wt{q}{}_A\,T^a q^A ~+~ 
\frac{1}{\sqrt{2}}\, \frac{\p\mc{W}_{\mc{N}=1}}{\p a^a} \Bigr|^2
~+~
\frac{g_1^2}{2}\, \Bigl| \wt{q}{}_A q^A ~+~ 
\sqrt{2}\, \frac{\p\mc{W}_{\mc{N}=1}}{\p\aU} \Bigr|^2
\\[3mm]
\notag
&\qquad\quad ~~+~~
2 \sum_{A=1}^{N_f} \Biggl\{  
\left| \left( \frac{1}{2}\,\aU ~+~ \frac{m_A}{\sqrt{2}} ~+~ a^a T^a \right) q^A \right|^2  ~+~
\\[3mm]
\notag
&\phantom{\qquad\quad ~~+~~ 2 \sum_{A=1}^{N_f} \Biggl\{  }
\left| \left( \frac{1}{2}\,\aU ~+~ \frac{m_A}{\sqrt{2}} ~+~ a^a T^a \right) \ov{\wt{q}}{}^A \right|^2  
\Biggr\}~,
\end{align}
where summation is implied over the repeated flavor indices $A$ (and over omitted color indices, too). 

Consider the case when we have one \textquote{extra} flavor, $N_f = N + 1$. Scalar potential \eqref{V} has a set of supersymmetric vacua, but in this paper we concentrate on a particular  vacuum where the maximal number of squarks equal to the rank of the gauge group $N$ condense. Up to a gauge transformation, the squark vacuum expectation values are given by%
%
%
\begin{eqnarray}
\langle q^{kA} \rangle ~=~ \langle \ov{\wt{q}}^{kA} \rangle& =& \frac{1}{\sqrt{2}}
\left(
\begin{array}{cccc|c}
  \sqrt{\xi_1} 	& 0  		& 0 				&0 								& 0 		\\
   0 			& \ddots   	& \vdots 			& \vdots 						& \vdots	\\  
  \vdots        &  \dots 	& \sqrt{\xi_{N-1}} 	&0 								& 0 		\\
 0 				& \dots 	& 0  				&  \sqrt{\xi_N}    				& 0 
\end{array}
\right)\,,
\label{qVEV}
\end{eqnarray}
where  we write quark fields as a rectangular matrices $N\times N_f$ and $\xi_P$ are defined as
\begin{equation}
	\xi_P ~~=~~ 2\left( \sqrt{\frac{2}{N}} \mu_1 \widehat{m} ~+~ \mu_2 (m_P - \widehat{m}) \right) ,
	\label{xi-general}
\end{equation}
\begin{equation}
	\widehat{m} ~~=~~ \frac{1}{N} \sum_{A=1}^{N} m_P .
\label{averagemass}
\end{equation}

If we define a scalar adjoint matrix as
\begin{equation}
	\Phi =  \frac{1}{2}\,a + T^a\, a^a \,,
	\label{Phidef}
\end{equation}
then the adjoint fields VEVs are given by
\beq
\langle \Phi\rangle = - \frac1{\sqrt{2}}
\left(
\begin{array}{ccc}
	m_1 & \ldots & 0 \\
	\ldots & \ldots & \ldots\\
	0 & \ldots & m_N\\
\end{array}
\right)\,.	
\label{avev}
\eeq
The  vacuum field (\ref{qVEV}) results in  the spontaneous
breaking of both gauge U$(N)$ and flavor SU($N$). However, in the equal mass limit $m_A \equiv m$, $A=1,...,N_f$ all
parameters $\xi$ become equal, $\xi_P\equiv  \xi$, $P=1,...,N$ and 
a diagonal global ${\rm SU}(N)_{C+F}$ survives, or, more exactly:
\begin{equation}
	{\rm U}(N)_{\rm gauge}\times {\rm SU}(N)_{\rm flavor}
	\to {\rm SU}(N)_{C+F} \times {\rm SU}(\wt{N})_{F} \times {\rm U}(1) \,.
\label{c+f}
\end{equation}
Thus, a color-flavor locking takes place in the vacuum. The presence of the  color-flavor ${\rm SU}(N)_{C+F}$ global symmetry is the reason for the formation of non-Abelian strings, see \cite{SYrev} for a review.

In the special case when
\begin{equation}
\mu_2 ~~=~~ \mu_1 \sqrt{2/N} ~~\equiv~~ \mu \ ,
\label{mu_condition}
\end{equation}
superpotential \eqref{none_superpotential_general} is simplified and becomes a single-trace operator
\begin{equation}
\mc{W}_{\mc{N}=1} ~=~ \mu \Tr (\Phi^2) \,.
\label{eq:none_superpotential_singletrace}
\end{equation}



\subsection{Mass spectrum  \label{sec:mass_spectrum}}

In this section we review the  mass spectrum of our bulk SQCD taking all quark masses equal, cf. 
\cite{VY,SYrev, BSYhet}. Due to squark condensation, the gauge bosons acquire masses%
\footnote{Here we assume for simplicity that  $\xi, \, \mu_1,\, \mu_2 $  are real}
\begin{equation}
\begin{aligned}
	m_{\ue} ~~=~~ g_1 \sqrt{\frac{N}{2}} \xi \,, \\
	m_{\sun} ~~=~~ g_2\sqrt{\xi} \,.	
\end{aligned}
\label{eq:gauge_mass}
\end{equation}

Scalar states masses are to be read off from the potential \eqref{V}. Expanding and diagonalizing the mass matrix one can find  $N^2 - 1$ real scalars with the masses $m_{\sun}$ and one scalar with the mass $m_{\ue}$. These are \none superpartners of SU$(N)$ and U(1)  gauge bosons. Other $N^2$ components are eaten by the Higgs mechanism. Another $2\times 2N^2 $ real scalars (  adjoint scalars $a^a$, $a$ and the half of squarks) become scalar components of  the following \none chiral multiplets: one  with mass
\begin{equation}
	\mUp ~~=~~ g_1 \sqrt{\frac{N}{2}\xi\lambda_1^+}~,
\end{equation}
and another one  with mass
\begin{equation}
	\mUm ~~=~~ g_1 \sqrt{\frac{N}{2}\xi\lambda_1^-}~.
\end{equation}
The remaining $ 2(N^2 - 1)$ chiral multiplets  have masses
\begin{equation}
	\mNp ~~=~~ g_2 \sqrt{\xi\lambda_2^+} ~,
\end{equation}
\begin{equation}
	\mNm ~~=~~ g_2 \sqrt{\xi\lambda_2^-} ~.
\end{equation}
Here  $ \lambda_i^\pm $ are roots of the quadratic equation \cite{VY,SYrev}
\begin{equation}
	\lambda_i^2  -  \lambda_i(2 + \omega^2_i)  +   1  =  0
\end{equation}
with
\begin{equation}
	\omega_1  =  \frac{g_1\mu_1}{\sqrt{\xi}}\,,\qquad
	\omega_2  =  \frac{g_2\mu_2}{\sqrt{\xi}}\,.
	\label{omega}
\end{equation}
Once $N_f>N$ apart from the above  massive scalars, we also have $4 N (N_f - N)$ scalars which come from the extra squark flavors $q^K$ and $\wt{q}_K$, $K=(N+1),...,N_f$. In the equal mass limit  these extra scalars are massless, and the theory enjoys a Higgs branch
\begin{equation}
{\cal H} = T^*\textrm{Gr}^{\mathbb{C}}(N_f, N) 
\label{HiggsbranchGr:3}
\end{equation}
of real dimension
\begin{equation}
	{\rm dim}{\cal H}= 4 N (N_f-N) \,.
	\label{dimH:3}
\end{equation}

In the large $\mu$ limit, states with masses $\mUp$ and $\mNp$ become heavy with masses $\sim g^2\mu$ and decouple. They correspond to the adjoint matter multiplets. Instead states  with masses $\mUm$ and $\mNm$ become light with masses $\sim \xi/\mu$. Scalar components of these multiplets are Higgs scalars. They develop VEVs
\eqref{qVEV}. 
In the opposite limit of small $\mu$ their masses are given by
\begin{equation}
\begin{aligned}
	\mUm &~~=~~ g_1 \sqrt{\frac{N}{2}\xi} \left(1 - \frac{g_1\mu_1}{2 \sqrt{\xi}} + \cdots\right)~,	\\
	\mNm &~~=~~ g_2 \sqrt{\xi} \left(1 - \frac{g_2\mu_2}{2 \sqrt{\xi}} + \cdots \right) ~.
\label{eq:m_scalar-small_mu}
\end{aligned}
\end{equation}
As we already mentioned  \ntwo supersymmetry is not broken in our theory to the leading order at small $\mu$
\cite{Hanany:1997hr,VY}. The leading order corresponds to sending parameters $\omega$ in \eqref{omega} to zero while keeping
FI parameter $\xi \sim \mu m$ fixed.
One can see that in the \ntwo limit  Higgs scalars are degenerate  with the gauge fields \footnote{They belong to the same long vector \ntwo supermultiplet \cite{VY}},  but become lighter as we switch on the $\mu$-deformation. 

The ratio of squares of Higgs and gauge boson masses $\beta$ is an important 
parameter in the theory of superconductivity. Type I superconductors correspond to $\beta< 1$, while type II
superconductors correspond to $\beta > 1$. BPS strings arise on the border at $\beta =1$. We see that in 
our theory both parameters $\beta$,
\beq
\beta_{U(1)} = \left(\frac{m^{-}_{U(1)}}{m_{U(1)}}\right)^2, 
\qquad \beta_{SU(N)} = \left(\frac{m^{-}_{SU(N)}}{m_{SU(N)}}\right)^2 ,
\label{betas}
\eeq
are less than unity, and thus our theory is in the  type I  superconducting phase at non-zero $\mu$.
This will turn out to be important later.

%
%

\section{Semilocal non-Abelian vortices  \label{sec:string-small_mu}}

In this section we study a vortex string solution in the equal quark mass limit. 
First we review  previous results \cite{ShYVinci} for the BPS semilocal non-Abelian vortex string and then consider  a small $\mu$-deformation. We   derive the world-sheet effective theory for the string moduli fields
in this case. For simplicity we consider the theory with one extra quark flavor, $N_f = N + 1$.

\subsection{BPS semilocal non-Abelian string}

We start by reviewing the semilocal non-Abelian string in the \ntwo limit \cite{ShYVinci}.
Once number of flavors exceed number of colors vortices  have no longer the conventional 
 exponentially small tails of the
profile functions. The presence of the Higgs branch and associated massless fields in the bulk makes them semilocal, see detail review of the Abelian case in \cite{AchVas}.
The semilocal  strings have a power fall-off at large distances from the string axis. For example, the  semilocal Abelian BPS string
interpolates between ANO string \cite{ANO} and two-dimensional  O(3) sigma-model instanton
uplifted to four dimensions (also known as the lump). For one extra flavor the semilocal string possesses
 two additional zero modes parametrized by the complex modulus $\rho$. The string's  transverse size 
is associated with $|\rho|$.
In the limit  $|\rho|\to 0$ in the Abelian case we recover the ANO string while
at $|\rho| \gg 1/m_{U(1)}$ it becomes a lump.

Consider an infinite static string stretched along the $x_3$ axis
 using the following ansatz:
\begin{equation}
q^{kA}=\bar{\tilde{q}}^{kA}=\frac1{\sqrt{2}}\vp^{kA}  ,
\label{q-ansatz}
\end{equation} 
\begin{equation}
	\vp ~~=~~ \bigg(\phi_{2}(r) + n\nbar (\phi_{1}(r)-\phi_{2}(r)) \,|\, n\,\phi_{3}(r) e^{- i \alpha}\bigg)
	\label{ansatz:vp:singular}
\end{equation}
for quarks, while the gauge fields are given by
\begin{equation}
	\begin{aligned}
		A_i^\text{SU($N$)} & ~~=~~ \varepsilon_{ij}\, \frac{x^j}{r^2}\, f_G(r)
		\lgr n\nbar ~-~ 1/N \rgr\,,
		\\[2mm]
		A_i^{\rm U(1)} & ~~=~~ \frac{2}{N}\varepsilon_{ij}\, \frac{x^j}{r^2}\, f(r)~.
	\end{aligned}
\label{string-solution}
\end{equation}
Index $i$ runs $i=1, 2$, all other components are zero; $\alpha$, $r$ are polar angle and radius in the 
$(x_1,x_2)$ plane respectively. The complex parameters $n^l, l=1,..,N$ obey the CP($N-1$) constrain $\ov{n}n = 1$. They parametrize the orientational zero modes of the non-Abelian string which appear due to the presence of the 
color-flavor group \eqref{c+f}, see \cite{SYrev} for a review.

The string profile functions entering \eqref{ansatz:vp:singular} and \eqref{string-solution} satisfy first order BPS equations. For  the case 
\begin{equation}
	\frac{g_1^2}{2} = \frac{g_2^2}{N} \equiv \frac{g^2}{N} 
\label{single_gauge}
\end{equation}
 the solution is particularly simple \cite{ShYVinci}. It is  is parametrized by a complex size modulus $\rho$:
\begin{equation}
\begin{aligned}
	\phi_1 ~~\approx~~&	\sqrt\xi\frac{r}{\sqrt{r^{2}+|\rho|^{2}}}									\,,		\\
	\phi_2 ~~\approx~~&	\sqrt\xi																	\,,		\\
	\phi_3 ~~=~~& \frac{\rho}{r}\phi_1 ~~\approx~~ \sqrt\xi\frac{\rho}{\sqrt{r^{2}+|\rho|^{2}}}		\,,		\\
	f ~~=~~ f_G	   ~~\approx~~&	\frac{|\rho^{2}|}{r^{2}+|\rho|^{2}} 										\,.
\end{aligned}
\label{eq:solution-semiloc-bps}
\end{equation}
This solution is valid in the limit $|\rho| \gg 1/(g_2\sqrt{\xi}|\rho|)$, i.e. when the scalar fields approach the vacuum manifold (Higgs branch). Tension of the BPS string is given by
\begin{equation}
	T_{BPS} = 2\pi\xi.
\label{eq:tension-semilocal}
\end{equation}

To obtain the low energy effective two dimensional theory living on the string world sheet, one should assume $n^P$ and  $\rho$ to be slowly varying functions of the transversal coordinates $t, z$, and substitute the solution \eqref{eq:solution-semiloc-bps} into the action \eqref{theory}. This procedure yields  the effective action
\begin{equation}
	S^{2d}_{SUSY} ~=~ \int d^2 x \left\{ 2\pi\xi \,	|\pt_k(\rho n_P)|^2	\,\,\ln{\frac{L}{|\rho|}}\,
		~+~ \frac{4\pi}{g^2} \Big[|\partial_{k}n_P |^2+(\nbar_P \pt_k n_P)^{2} \Big]
		\right\} ,
\label{eq:ws-bps}
\end{equation}
where the integration is carried over the coordinates $x_0, x_3$,
see the detailed derivation in \cite{ShYVinci}.
Here $k = 0, 3$, and $L$ is an infra-red (IR) cutoff introduced for the regularization of the logarithmic divergences
of orientational and size zero modes of the string. More exactly we introduce the  string of a large but finite length $L$.  This also regularize  the spread of string profile functions in the transverse plane 
\cite{Shifman:2006kd}.
The IR divergences arise due to the slow (power) fall-off  of the string profile functions associated with the presence of the Higgs branch \cite{Shifman:2006kd,ShYVinci}.

\subsection{Deformed world-sheet theory}

When we take into account higher order $\mu$-corrections, supersymmetry in the bulk reduces to \none, and 
as we already explained our theory becomes that of the type I superconductor, cf. \cite{VY}. The string  is no
longer BPS saturated. To mimic this we consider a simplified version of our theory with the bosonic action given by
\begin{multline}
	S_{\rm 0} =\int d^4x \Bigg\{ \frac1{4g_2^2}\left(F^{a}_{\mu\nu}\right)^{2}
		+ \frac1{4g_1^2}\left(F_{\mu\nu}\right)^{2}
		+   |\nabla_\mu \vp^A|^2														\\
		+	\lambda_N	\left(\bar{\vp}_A T^a \vp^A\right)^2
		+   \lambda_1 \left( | \vp^A|^2 - N\xi \right)^2
		\Bigg\}
	\, .
\end{multline}
This model depends on two parameters -- ratios of the squires of \ue and \sun Higgs and  gauge  boson masses given by
\begin{equation}
\begin{aligned}
	\beta_{U(1)} &= \frac{8\lambda_1}{g_1^2}\,, \\
	\beta_{SU(N)} &= \frac{2\lambda_N}{g_2^2}\,,
\end{aligned}
\label{eq:beta-qed}
\end{equation}
which we identify with $\beta$-parameters \eqref{betas} of our original theory .
The model above is a non-Abelian generalization  the one considered in \cite{GorskyShY_skyrmion}, where the  scalar QED was studied see also \cite{AchVas}.  

In \ntwo supersymmetric QCD parameters $\beta$ are exactly equal to one. In this case the  the Bogomol'nyi representation produces first order equations for the string profile functions. World sheet theory in this case is given by \eqref{eq:ws-bps}.

As we switch on $\mu$-corrections parameters $\beta$ are no longer equal to unity. Let us write the 
Bogomol'nyi representation for  the tension of the string
\begin{multline}
	T_\beta ~=~ \int d^2 x_{\perp}  \Bigg\{
			  \left[\frac1{\sqrt{2}g_2}F_{12}^{a} +	\frac{g_2}{\sqrt{2}}    \left(\bar{\vp}_A T^a \vp^A\right)\right]^2
			+ \left[\frac1{\sqrt{2}g_1}F_{12} +  \frac{g_1 }{2\sqrt{2}}	\left(|\vp^A|^2-N\xi \right)\right]^2			\\
			+  \left|\nabla_1 \,\vp^A +  i\nabla_2\, \vp^A\right|^2
			+ \frac{N}{2}\xi\,  F^{*}_3 																					\\
			+ \frac{g_2^2}{2} (\beta_{SU(N)} - 1) \left(\bar{\vp}_A T^a \vp^A\right)^2 
			+ \frac{g_1^2}{8} (\beta_{U(1)} - 1) \left( | \vp^A|^2 - N\xi \right)^2
			\Bigg\} \,,
			\label{bogomolny}
\end{multline}
where $\vec{x}_{\perp}$ represents the coordinates in the transverse plane.
Two extra terms written in the last line above appear. The Bogomol'nyi bound is no longer valid. But if the values $\beta_{U(1)}$ and $\beta_{SU(N)}$ only slightly differ from unity, then we can use the first order  equations to rewrite expressions in these extra terms as follows
\beq 
g_2^2 \left(\bar{\vp}_A T^a \vp^A\right) = - F_{12}^a, 
\qquad \frac{g_1^2}{2}\left( | \vp^A|^2 - N\xi \right) =- F_{12}.
\label{foe}
\eeq
In the case \eqref{single_gauge} we can use \eqref{eq:solution-semiloc-bps} to calculate the effective action.
Substituting \eqref{ansatz:vp:singular}, \eqref{string-solution}, \eqref{eq:solution-semiloc-bps} into 
\eqref{foe}, \eqref{bogomolny}  one arrives at the  deformed world-sheet theory,
\begin{multline}
 S^{2d}_{\beta} ~=~ \int d^2 x\left\{ 2\pi\xi \,	|\pt_k(\rho n_P)|^2	\,\,\ln{\frac{L}{|\rho|}}\, 
		+	\frac{4\pi}{g^2} \Big[|\partial_{k}n_P |^2+(\nbar_P \pt_k n_P)^{2} \Big]
		\right.
		\\
		\left.
		+ \frac{\beta - 1}{g^2} \, \frac{4\pi}{3|\rho|^2}\, +\cdots \, , 
	\right\} ,
\label{eq:ws-small_mu0}
\end{multline}
%
where now $\beta \equiv \beta_{U(1)} = \beta_{SU(N)}$ and the dots represent corrections in powers of $1/g^2\xi |\rho|^2$.

We see that for non-BPS string $\rho$ is no longer a modulus. It develops a potential proportional to
 the deviation of $\beta$ from unity. In particular, for type I superconductor ($\beta <1$) the size
$\rho$ tends to shrink, while for type II superconductor ($\beta >1$) the size
$\rho$ tends to expand making the vortex unstable, cf. \cite{AchVas}.

In our case, the value of $\beta$ is less then unity and is given
  by \eqref{eq:m_scalar-small_mu} at small $\mu$, namely
\begin{equation}
	\beta ~~=~~ 1 - \frac{g \mu}{\sqrt{\xi}} + \cdots\,.
\end{equation}
This gives the effective world sheet action on the string
\beqn
	&& S^{2d}_{\beta} ~=~ \int d^2 x\left\{ 2\pi\xi\,	|\pt_k(\rho n_P)|^2	\,\,\ln{\frac{L}{|\rho|}}\, 
		+	\frac{4\pi}{g^2} \Big[|\partial_{k}n_P |^2+(\nbar_P \pt_k n_P)^{2} \Big]
	\right.
	\nonumber\\
	&&
	\left.
	- \,4\pi \frac{\mu}{3 g\sqrt{\xi}} \, \frac{1}{|\rho|^2}\, + \cdots\, 
	\right\} .
\label{eq:ws-small_mu}
\eeqn

We see that the size of the semilocal string tends to shrink and at large $\mu$ we expect that 
the long-range tails of the string are not developed. The string becomes a  local non-Abelian string with only orientational moduli $n^l$, whose world sheet dynamics is described by CP$(N-1)$ model.

In fact we can argue on general grounds that as we turn on $\mu$ and make it large  the semilocal string  become unstable. The semilocal string solution
\eqref{eq:solution-semiloc-bps} is ''made'' of massless fields associated with the Higgs branch of the theory.
As we already mentioned say, in the Abelian case this solution  correspond to the instanton of the two dimensional
O(3) sigma model uplifted to four dimensions. The instanton is  essentially  a BPS solution and therefore it is natural to expect that it becomes unstable once we increase $\mu$ breaking the world sheet  supersymmetry.

In particular, as we see from Bogomol'ny representation \eqref{bogomolny} extra terms arising at $\beta <1$
reduce the tension of the string. This is forbidden for BPS lump (uplifted instanton) since its tension
is exactly determined by the central charge and given by $2\pi \xi$, see \eqref{eq:tension-semilocal}.
As we increase $\mu$ the string is no longer BPS, $\rho$ develops instability and shrinks leading at large $\mu$
to much lower tension, see \eqref{eq:T_local} below.

%
%

\section{Summary of  results \label{sec:results}}

In this paper we studied what happens to the non-Abelian semilocal string in \ntwo supersymmetric QCD as we switch on the $\mu$ deformation and go to the large $\mu$ limit. We showed that the size modulus $\rho$ develops a potential and eventually decouples as the theory flows to the \none SQCD at large $\mu$. 
Note that the Higgs brunch is still there, just the string is no longer ''made'' of massless fields,
so the long-range ''tails'' of the string disappear.

Thus,  the semilocal string degenerates into the local one. Non-Abelian local strings  in the large $\mu$ limit
of \none SQCD  were studied in \cite{YIevlevN=1}, and now we see that  results of this  paper can be directly applied to our case $N_f >N$ as well.
Below we briefly summarize these results.

In the large $\mu$ limit, the string tension is logarithmically suppressed \cite{YIevlevN=1}, 
\begin{equation}
	T_{local} = \frac{4\pi|\xi|}{\ln \displaystyle\frac{g^2|\mu|}{|m|}}  \,.
\label{eq:T_local}
\end{equation}
This should be contrasted with the BPS formula \eqref{eq:tension-semilocal} valid to 
the leading order at small $\mu$.

As usual the world sheet  theory contains  translational moduli but they decouple from the orientational sector.  The orientational sector is described by  CP$(N-1)$ model with the action
\begin{equation}
	S^{(1+1)}=    \int d t\, dz \,  
			\Big\{ 
				\gamma\, \left[ (\pt_{k}\, \bar{n}\,	\pt_{k}\, n) + (\bar{n}\,\pt_{k}\, n)^2 \right]
				~+~  V_{1+1}
			\Big\}\,.
\label{cp}
\end{equation}
Note, that orientational fermionic zero modes are all lifted \cite{YIevlevN=1} and do not enter 
the low energy world sheet theory. The above world sheet theory is purely bosonic.

Here two dimensional inverse coupling constant $\gamma$ is large, given by
\begin{equation}
	\gamma  \sim \frac{|\mu|}{|m|}\,\frac{1}{\ln^2 \frac{g^2|\mu|}{|m|}}. 
\label{beta}
\end{equation}
At the quantum level CP$(N-1)$ model is asymptotically free, so the coupling $\gamma$ runs and at the energy  $E$ is given by
\beq
2\pi\gamma (E)= N \log{\left(\frac{E}{\Lambda_{CP}}\right)},
\eeq
where the scale of 
the  world sheet theory is given by
\begin{equation}
	\Lambda_{CP} \approx \sqrt{\xi} \exp{\left(- {\rm const}\,\frac{|\mu|}{|m|}\frac{1}{\ln^2 \frac{g^2|\mu|}{|m|}}\right)}.
\label{LambdaCP}
\end{equation}
We see that the scale $\Lambda_{CP}$ of CP$(N-1)$ model above  is exponentially small, so the world sheet theory
is weakly coupled in a wide region of energies $\gg \Lambda_{CP}$. This should be  contrasted to non-Abelian string in \ntwo QCD where world sheet theory has a scale $\Lambda_{CP}$ equal to scale 
$\Lambda_{{\cal N}=2}$ of the bulk SQCD \cite{SYrev}.

In the case when the quark masses entering the Lagrangian \eqref{theory} are non-identical, a potential for $n^P$ is generated. In the simplest case when all quark masses are positive, this potential is given by \cite{YIevlevN=1}
\begin{equation}
	 V_{1+1} ~~\approx~~ \frac{8\pi |\mu|}{\ln \frac{g^2|\mu|}{|m|}}     \sum_{P=1}^{N} 
			 m_P |n^P|^2 \,.
\label{V1+1}
\end{equation} 
The potential \eqref{V1+1} has only one minimum and one maximum at generic $\Delta m_{AB}$. Other $(N-2)$ extreme points are saddle points. For equal quark masses this potential reduces to the constant equal to the tension of the string \eqref{eq:T_local}.

Since our four-dimensional theory is in the Higgs phase for squarks, 't Hooft-Polyakov monopoles present in the  theory
in the \ntwo limit of small $\mu$ are confined by non-Abelian strings and serve as junctions of two distinct strings \cite{T,SYmon,HT2}. In the 
effective world sheet theory on the non-Abelian string they are seen as kinks interpolating between 
different vacua of CP$(N-1)$ model, see \cite{SYrev} for a review.

In the large $\mu$ limit adjoint fields decouple. Therefore we could expect quasiclassically that the confined monopoles disappear in this limit. This indeed happen for non-equal quark masses. If quark mass differences  are non-zero, a potential  \eqref{V1+1} is generated.
It does not have multiple local minima, therefore kinks (confined monopoles of the bulk theory) 
become unstable and disappear.

However, in the equal quark mass case the potential \eqref{V1+1} is absent and the bosonic CP$(N-1)$ model supports kinks. Thus, in this case confined monopoles do survive the large $\mu$ limit \cite{YIevlevN=1}. The monopoles are represented by kinks in the effective CP$(N-1)$ model on the non-Abelian string, see \cite{SYrev} for a detail review.

%
%

\section*{Acknowledgments}

The authors are grateful to Mikhail Shifman for very
useful and stimulating discussions. 
The work of E. I. was funded by RFBR according to the research projects No. 18-32-00015 and No. 18-02-00048.
The work of A.Y. was  supported by William I. Fine Theoretical Physics Institute  at the  University 
of Minnesota and  by Russian Foundation for Basic Research Grant No. 18-02-00048.

%
%

%
%



\end{document}